\documentclass[aps,prl,a4paper,superscriptaddress,floatfix,preprintnumbers,twocolumn]{revtex4-1}

\usepackage[utf8]{inputenc}
\usepackage[T1]{fontenc}
\usepackage{amsmath,amsthm,amssymb}

\usepackage{verbatim}
\usepackage{natbib}
\usepackage{bibunits} 
\usepackage{bbm}

\usepackage{graphicx} 
\usepackage{tikz}
\usetikzlibrary{decorations.pathmorphing}
\usetikzlibrary{arrows}
\usetikzlibrary{matrix}
\usepackage[normalem]{ulem}
\definecolor{gray}{rgb}{.6,.6,.6}

\usepackage{hyperref}
\usepackage{verbatim}

\newcommand{\id}{\ensuremath{\mathbbm{1}}} 
\newcommand{\ket}[1]{\ensuremath{\left| #1\right>}}

\newcommand{\ketbra}[2]{\ensuremath{\vert\!\!\;#1\rangle \! \langle #2\vert}}
\newcommand{\proj}[1]{\ensuremath{\ketbra{#1}{#1}}}
\newcommand{\tr}{\mathrm{tr}}

\def\cB{{\cal B}}
\def\cD{{\cal D}}

\def\cH{{\cal H}}
\def\cP{{\cal P}}
\def\cR{{\cal R}}
\def\cS{{\cal S}}
\def\cT{{\cal T}}
\def\cU{{\cal U}}

\def\cW{{\cal W}}
\newcommand{\Eqref}[1]{Eq.~(\ref{#1})}                                

\newcommand{\Ineqref}[1]{Ineq.~(\ref{#1})}

\newcommand{\Leref}[1]{Lemma~\ref{#1}}                  
\newcommand{\Thref}[1]{Theorem~\ref{#1}}
 
\newcommand{\mr}[1]{\mathrm{#1}}
\def\Proof{\textsc{Proof}}
\def\assign{\mathrel{\raise.095ex\hbox{\ensuremath{\mathrm{:}}}\mkern-4.2mu=}}

\newtheorem{theorem}{Theorem}

\newtheorem{lemma}{Lemma}
\newtheorem{example}{Example}
\newtheorem{definition}{Definition}

\newcommand{\SO}{\mathrm{SO}} 
\newcommand{\Inv}[1]{\frac{1}{#1}} 

\newcommand{\OSEp}{O_{SE'}} 
\newcommand{\OSE}{O_{SE}} 

\newcommand{\affiliationMPQ}{\affiliation{Max-Planck-Institut f\"{u}r Quantenoptik, 85748 Garching, Germany}}
\newcommand{\affiliationIKER}{\affiliation{IKERBASQUE, Basque Foundation for Science, 48013 Bilbao, Spain}}
\newcommand{\affiliationDIPC}{\affiliation{Donostia International Physics Center, 20018 San Sebasti\'an, Spain}}
\newcommand{\affiliationAarhus}{\affiliation{Department of Physics and Astronomy, Aarhus University, 8000 Aarhus, Denmark}}
\newcommand{\affiliationETH}{\affiliation{Institute for Theoretical Physics, ETH Zürich, 8093 Zürich, Switzerland}}
\begin{document}

\author{Eli\v{s}ka Greplov\'a}\affiliationMPQ\affiliationAarhus\affiliationETH
\author{G\'eza Giedke}\affiliationMPQ\affiliationIKER\affiliationDIPC

\date{\today}

\title{Degradability of Fermionic Gaussian Channels}
\begin{abstract}
  We study the degradability of fermionic Gaussian channels.
  Fermionic quantum channels are a central building block of
    quantum information processing with fermions, and the family of
    Gaussian channels, in particular, is relevant in the emerging field
    of electron quantum optics and its applications for quantum
    information. Degradable channels are of particular interest since
    they have a simple formula that characterizes their quantum capacity. We derive a simple standard
  form for fermionic Gaussian channels.  This allows us to fully
  characterize all degradable $n$-mode fermionic Gaussian channels. In particular, we show that the only degradable such channels
correspond to the attenuation or amplitude-damping channel
for qubits.
\end{abstract}

\maketitle

\begin{bibunit}[apsrev4-1] 
The transmission of quantum states in space and time is a fundamental
physical process, described by quantum
channels \cite{Wolf2012}. Therefore the properties of quantum channels and their capacity to transmit classical or quantum
information \cite{Wilde2017} is central to quantum information processing.  Channel capacities are difficult to
compute since, in general, they require an optimization over entangled
inputs to many channels in parallel \cite{Sho02,Hastings2009} and are
only known for few channels.

These complications do not arise for the quantum
capacity of \emph{degradable} channels
\cite{DeSh05,CRS08}, which can be expressed as a simple formula (which also
equals their private capacity \cite{Smi08}). Their characteristic property is that the state of the environment
can be reproduced from the channel output by applying another quantum
channel.  The notion has been generalized to weak \cite{CGH06}, conjugate
\cite{Bradler2010},  and approximate \cite{SSR14} degradability,
maintaining some of its useful properties.

The most natural information carrier in solid state systems are
electrons (quantum dot electrons \cite{LoDi98,HF+17},
or, more recently, Majorana fermions in quantum wires \cite{SFM15,LBK+18,Mourik1003}), i.e., fermions, whose
anti-commutation and superselection rules necessitate the refinement
of central concepts of quantum information theory such as entanglement
\cite{ESBL02,BoRe04b,BCW07,FLB13,EiZi15,SSR17,EEZ18}.  Impressive
experimental advances (e.g., edge channels
\cite{HOS+Schoenenberger99,BF+Feve14}, moving quantum dots
\cite{MK+Ritchie11,HT+Meunier11,Bertrand2016,Ford2017}, quantum dot arrays
\cite{Fujita2017}), demonstrate that electrons can be cleanly and
individually transported in well-controlled semiconductor systems,
providing fermionic quantum channels over sample-scale
distances. These may serve for on-chip information transfer, e.g.,
between different registers of a quantum processor. This progress motivates the
study of \emph{fermionic quantum channels}
\cite{Bravyi2005,SS+Lewenstein05,BJJ11,MCT13}, which are also useful to
describe storage (transmission in time) of quantum information using
fermionic systems.
 
Fermionic quantum channels have mostly been studied for
non-interacting fermions, leading to the notion of fermionic Gaussian
channels (FGC) \cite{Brav05} (also known as ``quasifree'' channels
\cite{Eva79,DFP08} or ``fermionic linear optics''
\cite{Kni01,Brav05,DiTe05,JoMi08,MCT13}). These works emphasize the analogy
with the case of Gaussian bosons, which are a very fruitful model for
optical quantum information processing \cite{Weedbrook2012, Serafini2017}. Here we
exploit fermionic phase space methods to
analyze the degradability of FGCs. We derive a simple standard form
which simplifies further analysis. With the
phase-space characterization of quantum channels \cite{Brav05} in
  this form we give a full characterization of all degradable $n\to n$
mode FGCs and show that there is only one family of such channels, the
single-mode attenuation channel (see Theorem \ref{th:th1}).

We consider free fermions with $n$-dimensional
one-particle Hilbert space $\cH$ (``$n$ modes'')
\cite{Bravyi2005,Der06}, described by $2n$ Hermitian
operators $c_k, k=1,\dots,2n$, satisfying
$\{c_k,c_l\}=2\delta_{kl}$ and associated annihilation
$a_j=(c_{2j-1}-ic_{2j})/2$ and creation operators $a^\dag_j,
j=1,\dots,n$.

\emph{Fermionic Gaussian states (FGS)} are those states for which
Wick's theorem holds \cite{CoDr06} (all cumulants are zero). They are
fully described by the $2n\times 2n$ \emph{covariance matrix} (CM)
\cite{Brav05,DFP08} defined as
\begin{equation}\label{eq:CM}
  \gamma_{kl} =
  \frac{i}{2}\textup{tr}\left(\rho[c_k,c_l]\right). 
  \end{equation}
The matrix $\gamma$ is real and antisymmetric. We frequently use that any such matrix can be brought to the form \cite[p.~18]{gantmakher2000theory}
$\Lambda=\oplus_{j=1}^n \lambda_jJ$, with $J=\left( \begin{array}{cc} 0 & -1\\ 1 &0
  \end{array} \right)$ by a special orthogonal transformation: there exist $\lambda_j\in[-1,1]$ and
$O\in\SO(2n)$ such that $\gamma=O\Lambda O^T$.  The CM 
$\Lambda$ describes $n$ modes in Gibbs states for the Hamiltonian $a_j^\dag a_j$. The 
Gaussian state is pure if and only if (iff) all $\lambda_j=\pm1$, or, equivalently, iff   
$\gamma^2=-\mathbbm{1}$ holds.

If we consider a bipartite system of $n+m$ modes, then simplification
of $\gamma$ under \emph{local} operations $\SO(2n)\oplus\SO(2m)$ is of
interest. Any pure state $\Gamma$ can be brought into
the Schmidt-form \cite{BoRe04b} with CM
\begin{align}
\label{eq:BRform}
\left(\begin{array}{cc}
\Gamma_{11}&\Gamma_{12}\\ \Gamma_{21}&\Gamma_{22}
\end{array}\right)&\equiv
\left( \begin{array}{cc}
J_{2l}\oplus\Lambda & \left(\begin{array}{cc}0_{2m\times
  2l}&K\end{array}\right)^T \\
\left(\begin{array}{cc}0_{2m\times
  2l}&-K\end{array}\right) &  \Lambda
\end{array} \right)
\end{align} 
by such local operations.  Here $n=l+m$ and $J_{2s} = \oplus_{j=1}^{s}J$,
$\Lambda=\oplus_{j=1}^{m} \lambda_j J$, and $K = \oplus_{j=1}^{m} \kappa_j \sigma_x$,
with $\lambda_j^2+\kappa_j^2=1$ (the parameters $\kappa_j$ specify the amount
of entanglement between the two parties).

Now let us turn to \emph{Fermionic Gaussian channels}.  We consider
quantum channels (trace-preserving completely positive maps) that act
on a finite set of $n$ fermionic modes and map Gaussian states to
Gaussian states. As discussed in \cite{Brav05} they are fully
characterized by how they transform the $2n\times 2n$ covariance
matrix $\gamma$ of the input state. An $n\to m$ mode FGC $\cT$ is
defined by a $2m\times 2n$ matrix $A$ and an antisymmetric $2m\times
2m$ matrix $B$ as
\begin{equation}
  \label{eq:FGC}
  \cT\equiv \cT_{(A;B)}:\gamma \mapsto A\gamma A^T+B.
\end{equation}

Equivalently, a channel $\cT_{(A;B)}$ can be characterized via its
Choi-Jamiolkowski (CJ) state \cite{CRS08}, which is given by the state
obtained if the channel acts on the first half of a maximally
entangled state. For Gaussian channels, the CJ state is Gaussian with
CM $M_{(A;B)}=\left( \begin{array}{cc} B & A\\ -A^T &0
  \end{array} \right)$, since the maximally entangled state of $2n$ fermionic
modes can be chosen Gaussian [$l=0$ and $\lambda_j=0, \kappa_j=1, \forall j$ in
\Eqref{eq:BRform})].  This yields a practical necessary and sufficient
criterion for $(A;B)$ to define a valid quantum channel \cite{Brav05}:
$\cT_{(A;B)}$ describes a valid FGC iff the corresponding CJ-CM is a
valid CM, i.e., iff $\id+iM_{(A;B)}\geq0$, which is readily seen (see
the Supplemental Material~\cite[Lemma S1]{SuppInfo}) to be the case iff
\begin{equation}
  \label{eq:CPcond}
  \id-iB-AA^T\geq 0.
\end{equation}
This implies that $B$ is a valid CM and that $\id-AA^T\geq0$ for FGCs
and that the kernel of $B$ contains that of $\id-AA^T$: $\ker
B\supseteq\ker(\id-AA^T)$. Thus, the singular values of $A$
must be $\leq1$ and $B$ must vanish on the unit eigenspace of $AA^T$
(the perfectly transmitted modes). This ensures that $B' =(\id-AA^T)^{-1/2}B(\id-AA^T)^{-1/2}$ is well-defined
and a CM (the inverse denotes the Moore-Penrose pseudo-inverse
\cite{HJ87} if $\mathbbm{1}-AA^T$ has a kernel).

\begin{figure}[h]
  \centering
  \includegraphics[width=75mm]{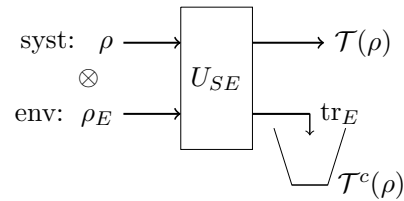}
  \caption{Channel and complementary channel. For \emph{pure}
    environmental state $\rho_E$ the dilation $U_E$ and the
    complementary channel are unique (up to isometries).}
  \label{fig:Qchan}
\end{figure}

Every quantum channel $\cT$ can be represented as a unitary
acting on the system and an initially factorized environment prepared in some state $\rho_E$, see Fig. \ref{fig:Qchan}. 
Thus $\cT$ comes with a second channel, which describes what
is leaked into the environment. For
pure $\rho_E$, this channel is unique up to isometries and is called
the \emph{complementary channel} (of $\cT$), denoted by $\cT^c$.  For
our purposes, all these complementary
channels are equivalent (see \cite[Lemma S3]{SuppInfo}).

The relation between $\cT$ and $\cT^c$ has important consequences for
certain capacities of $\cT$ \cite{DeSh05,Smi08,CRS08}. E.g., if it
were possible to obtain, for every input $\rho$, the channel output
$\cT(\rho)$ by suitably post-processing the output of $\cT^c$ then the
channel $\cT$ has vanishing quantum capacity, since any non-zero
capacity would contradict the no-cloning principle \cite{CRS08}.  Such
channels for which there exists a completely positive (CP) map $\cP$ such
that $\cP(\cT^c(\rho))=\cT(\rho)$ are called \emph{antidegradable}
\cite{CaGi06}.

A more subtle consequence holds if there exists a quantum channel
$\cW$ such that the concatenation of $\cW$ and $\cT$ is equal to the
complementary channel $\cT^c$. Such channels are called
\emph{degradable} and have quantum capacity that can be characterized by a simple formula \cite{DeSh05}.
Degradable and antidegradable channels (and their ``conjugate''
relatives \cite{Bradler2010,Wat12}) are at present the only ones for
which a good understanding of their quantum capacity can be claimed.
As we shall see, the simple structure of fermionic Gaussian channels
allows a straightforward answer to the question which $n\to n$
channels are degradable. This is in contrast to the available
characterization for bosons, where the full degradability description
is restricted to the one-mode case \cite{CaGi06,WPG07} or to the notion of
weak-degradability \cite{CEGH08}.  Note that while degradability can be proven
by constructing a degrading channel \cite{CRS08}, to show that a
channel is not degradable requires to show that \emph{none} of the
(possibly many, non-equivalent \cite{Bra15}) degrading
maps is CP. In the case of interest here, we can show
that the map is effectively unique. Our main result is summarized in
the following theorem:
\begin{theorem}
\label{th:th1}
All  degradable $n\to n$ fermionic Gaussian channels act on the
  covariance matrix $\gamma$ as
      \begin{equation}
        \label{eq:DegChannelForm}
        \cT_p:\gamma\mapsto (1-p)\gamma+pJ_{2n}, 
      \end{equation}
up to unitary pre- and post-processing or are a direct sum of such channels. Here,
 $0\leq p\leq \frac{1}{2}$ and $J_{2n}=-i\oplus_{j=1}^n\sigma_y$.
\end{theorem}
The proof proceeds in three steps: first we observe that concatenating a
channel with unitary channels does not affect (anti)degradability. This allows to simplify
the further discussion by focusing on FGCs in \emph{standard form} in
which the matrix $A$ is diagonal with descendingly ordered, positive
eigenvalues. This form can always be reached by concatenating the
channel with Gaussian unitaries that effect the singular value
decomposition of $A=O_1DO_2$ (see \cite[Lemma S6]{SuppInfo}).
Additionally,  FGCs that act independently on two
subsets of modes, i.e., $\cT_{(A;B)}$ with $A=A_1\oplus A_2$ and
$B=B_1\oplus B_2$, are (anti)degradable iff both $\cT_{(A_i;B_i)}$ are
(see \cite[Lemma S7]{SuppInfo}). Furthermore, a degrading map (CP
  or not) can only exist if $A$ is invertible and in this case the FGC in
  question is itself \emph{invertible}. Then the degrading map is
  unique (see \cite{Bra15}, \cite[Lemma S4]{SuppInfo})
   and the FGC is degradable iff the degrading map is
  CP. 
We then show that it is necessary that 
$D\geq1/\sqrt{2}$ and that the channel has a small Choi rank
\cite{CRS08} (its minimal dilation requires no more
than $n$ modes). Finally, we prove that such channels cannot be
degradable unless $D$ is a direct sum of
(even-dimensional) terms $\propto\id$
($D=\oplus\alpha_k\id_{2n_k}$ for $\alpha_k>0$) and the environmental state CM
$\gamma_p$ a direct sum of pure CMs $\oplus_k \gamma_{p,k}$ (same
partition as $D$). 

Let us now construct the standard form and the corresponding degrading map. Without loss of generality we consider an $n\to m$ FGC $\cT_{(A;B)}$
with 
\begin{equation}
  \label{eq:SVDform}
  A= \left( \begin{array}{cc} D & 0_{2(n-m)}\end{array} \right)
 \,\,\, \mr{or} \,\,\,
A = \left( \begin{array}{c} D\\ 0_{2(m-n)} \end{array} \right)
\end{equation}
depending on
whether $m\leq n$ or $m\geq n$. Here $1\geq D\geq0$ is a
square matrix with dimension $2\cdot\mr{min}\left\{ n,m \right\}$. 

If $D$ has $L\geq2$ eigenvalues equal to $1$ it implies that
$\left\lfloor L/2\right\rfloor$ modes are transmitted perfectly. As
$B$ then vanishes on all those modes, the channel is a particular case
of a Gaussian product channel $(A;B) = (\id\oplus A_2;0\oplus B_2)$
and thus it is degradable iff $\cT_{(A_2;B_2)}$ is, where $A_2$ now
has at most one singular value equal to 1 and it suffices to consider such channels in our proof.

First, we need the complementary channel to
$\cT_{A;B}$ in order to express degradability in terms of $A$ and
$B$. To this end, it is useful to find a unitary dilation of
$\cT_{A;B}$.  From \Eqref{eq:CPcond}, we see that for FGCs
$AA^T\leq1$ and $\mr{ker}(\id-AA^T)\subseteq\mr{ker}(B)$, and
$B'\assign(\id-AA^T)^{-\frac{1}{2}}B(\id-AA^T)^{-\frac{1}{2}}$ is a
valid CM.  Then it is easy to check that the $n\to m$ FGC $\cT_{A;B}$
can be obtained by a fermionic Gaussian unitary
represented by $\OSEp\in \mathrm{SO}(2n+2m)$,
\begin{equation}
  \label{eq:OSEp}
  \OSEp = \left( \begin{array}{cc} 
A & \sqrt{\id_{2m}-AA^T} \\
-\sqrt{\id_{2n}-A^TA} & A^T
\end{array}\right),
\end{equation}
acting on the system and an $m$-mode environment in the Gaussian state with CM $B'$. To obtain the
complementary channel, however, the environment should be
\emph{pure}. 
Let $l\leq n$ denote the number of pure modes of $B'$,
i.e., $B'=O(J_{2l}\oplus L)O^T$ for $0\leq\lambda_j<1$, 
$L=\oplus_{j=1}^{m-l}\lambda_j J_2$, and $O\in\SO(2m)$. Then 
\begin{equation}
  \label{eq:gammaE}
  \gamma_E = [O\oplus\id_{2(m-l)}]
\left( \begin{array}{cc|c} J_{2l} & \\
& L & K\\
\hline
& -K & L
\end{array} \right)[O^T\oplus\id_{2(m-l)}],
\end{equation}
where $K=\oplus_{j=1}^{m-l} \kappa_j\sigma_x$ and $\lambda_j^2+\kappa_j^2=1$ is
a purification of $B' $ and
$\cT_{A;B}$ can be obtained by coupling with
\begin{equation}
  \label{eq:OSE}
  \OSE = \OSEp\oplus \id_{2(m-l)}
\end{equation}
to the $2m-l$-mode pure environment in state $\gamma_E$. 

There are other physical representations of $\cT_{A;B}$ with pure
environment $\gamma_E'$ but they are all related isometrically to each
other \cite{Wolf2012} and are all equivalent for our purposes (see \cite[Lemma S3]{SuppInfo}).

Using this representation of $\cT_{A;B}$ we can read off its complementary channel 
$\cT^c_{A;B}$. It is the $n\to n+m-l$ map given by
  \begin{equation}\label{eq:CompChan}
    \cT^c_{(A;B)}\equiv\cT_{(A_c;B_c)}:\gamma\mapsto A_c\gamma A_c^T + B_c,
  \end{equation}
where
\[
A_c = \left( \begin{array}{c} \sqrt{\id-A^TA}\\0
\end{array} \right);\,\,\,\, B_c = (A^T\oplus\id)\gamma_E(A\oplus\id)
\]
with $\gamma_E$ as in \Eqref{eq:gammaE}.

The question of the degradability of the FGC $\cT_{A;B}$ is then, simply, if
there exists an $m\to n+m-l$ FGC $\cT_{\tilde{A};\tilde{B}}$ such that 
$\cT_{\tilde{A};\tilde{B}}\circ\cT_{A;B}=\cT_{A_c;B_c}$.
The degrading map  follows directly from
$\cT_{A;B}$ and $\cT_{A_c;B_c}$. 
The map only exists if $A$ has no kernel and is then 
given by  $\gamma\mapsto \tilde{A}\gamma \tilde{A}^T+\tilde{B}$ with
\begin{align}
  \label{eq:Ad}
  \tilde{A}&= 
\left( \begin{array}{c} \sqrt{\id-A^TA}\\ 0
\end{array} \right) A^{-1} = 
\left( \begin{array}{c} A^{-1}\sqrt{\id-AA^T}\\ 0
\end{array} \right),  \\
  \label{eq:Bd}
\tilde{B}&=(A^T\oplus\id)\gamma_E(A\oplus\id) -\\
&{}\,\,\,\,\,\,\,- \left[(A^{-1}-A^T)\oplus 0\right]\gamma_E\left[(A^{-T}-A)\oplus0\right]\nonumber.
\end{align}
Using \Eqref{eq:CPcond} we see that $\cT_{(\tilde{A};\tilde{B})}$ is
CP iff
\begin{equation}
  \label{eq:poscond2}
  \tilde{M}\equiv\id -\tilde{A}\tilde{A}^T-i\tilde{B}\geq0.
\end{equation}
We have thus constructed a fermionic Gaussian degrading map for
  a given FGC whenever it exists. 
It is then straightforward to check if it is CP via
\Eqref{eq:poscond2} and we now characterize all FGCs 
$(A;B)$ for which this is the case. Note that it is sufficient to check the
properties of the map $\cT_{(\tilde{A};\tilde{B})}$ 
since the channel $\cT_{(A;B)}$ is invertible for invertible $A$ (cf. \cite{SuppInfo}), 
and the degrading map is unique in this case \cite{Bra15}.
One may wonder whether our purely Gaussian discussion allows for 
possible non-Gaussian degrading maps. But since the Gaussian states span the
space of all fermionic density matrices (see \cite[Lemma S5]{SuppInfo}), 
$\cT_{(A;B)}$ is indeed invertible as a linear map on the space of
fermionic density matrices and the Gaussian
degrading map Eqs.~(\ref{eq:Ad},\ref{eq:Bd}) is unique.

First, we claim that \emph{for a $n\to n$ FGC to be degradable it is
    necessary that its Choi rank is $\leq n$ modes}.  Assuming standard form
$A=D$, re-expressing $\tilde{M}$ in \Eqref{eq:poscond2} in terms of
$D$ and the pure $2n-l$ mode environmental state $\gamma_E$ of a
minimal dilation we obtain after repeated application of 
the Schur complement to check positivity of a block matrix (see \cite[Lemma S1]{SuppInfo})
the inequality
\begin{equation}
  \label{eq:poscond4}
  2D^{-2}-D^{-4}-
\left[ O(0_{2l}\oplus\id_{2(n-l)})O^T \right]\geq 0,
\end{equation}
as a necessary condition for degradability, where $O\in\SO(2n)$
depends on $\gamma_E$  (for the details see \cite{SuppInfo}).  
This inequality cannot be fulfilled unless
$l=n$, i.e., the environment is no larger than the system. 
To see this, we use a condition on the eigenvalues of two
Hermitian matrices and their sum implied by Horn's conjecture,
\cite{Horn62,KnTa01}: Let $\lambda_i,\mu_j,\nu_k$ denote the descendingly
ordered eigenvalues of the Hermitian matrices $X,Y,X+Y$, respectively.  Then
we have \cite{KnTa01}
\begin{equation}
  \label{eq:Horn1}
  \nu_k\leq\lambda_i+\mu_j\,\,\,\forall i+j=k+1.
\end{equation}
We take $X=2D^{-2}-D^{-4}$ and
$Y=-O(0_{2l}\oplus\id_{2(n-l)})O^T$ and pick
$j=2l+1$. Then $\mu_j=-1$ and for all $i>1$ we have
\[
  \nu_{2l+i}\leq\lambda_i+\mu_{2l+1}=\frac{2}{d_i^2}-\frac{1}{d_i^4}-1\,\,\,\forall i=1,\dots,2(n-l).
\]
Unless $n-l=0$ (pure environment) we can take $i=2$ which means that
$d_i<1$ (since in standard form we have at most one singular value
  of 1) in which case the expression on the RHS is negative.

Let us now focus on $n\to n$ FGC with a $n$-mode
environment. To complete the proof of \Thref{th:th1}, we show 
\begin{lemma}[Only constant-loss channels are degradable]
  A $n\to n$ FGC $\cT_{(D;B)}$ in standard form with 
  Choi rank of $\leq n$ modes is degradable iff $D=\oplus_j (d_j\id_{2n_j})$, $d_j\geq1/\sqrt{2}$,
  and $B=\oplus_j B_j$. 
\end{lemma}
\textsc{Proof}: 
Following analogous arguments to the construction above (in particular, using
that $B=\sqrt{\id-D^{-2}}\gamma_p \sqrt{\id-D^{-2}}$ for a $n$-mode pure-state CM
$\gamma_p$)
 the degradability condition \eqref{eq:poscond2} becomes
\begin{equation}\label{ineq:le2}
2\id - D^{-2}-i\left[  D\gamma_p D-\left(\Inv{D}-D\right)\gamma_p\left(\Inv{D}-D\right)\right]\geq0.
\end{equation}

We show now that this only holds if $D\geq1/\sqrt{2}$ and if
$(\gamma_p)_{kl}=0$ whenever $d_k\not= d_l$, i.e., 
if $\gamma_p$ is a direct sum of pure
CMs $\oplus_m
\gamma_{p,m}$ and $D$ a corresponding direct sum of terms proportional
to $\id$.  

We already saw that $D\geq1/\sqrt{2}$ is necessary for degradability.
If there are one or more eigenvalues $d_i=1/\sqrt{2}$, then the real part
of \eqref{ineq:le2} has a kernel and the
inequality can only hold if $\gamma_{ij}=\gamma_{ji}=0$ for all $j$ such that
$d_j\not=1/\sqrt{2}$. I.e., a channel with some $d_j=1/\sqrt{2}$ can
only be degradable if $D=D'\oplus \frac{1}{\sqrt{2}}\id$ and
$\gamma=\gamma_1\oplus\gamma_2$ in accordance with \Thref{th:th1} (by
purity and antisymmetry, both blocks have to have even dimension).

We now assume $D>1/\sqrt{2}$. Multiplying \Eqref{ineq:le2} by
$\frac{1}{\sqrt{2-D^{-2}}}$ from left and right, the imaginary part
becomes $\gamma_p+R$, where 
\begin{equation}
  \label{eq:RDg}
  R = -\gamma_p+\Inv{W}\left[\gamma_p-
 D^2\gamma_p-\gamma_pD^2 \right]\Inv{W}
\end{equation}
with $W=\sqrt{2D^2-1}$.

Since $\gamma_p$ is pure, $i\gamma_p$ has spectrum $\left\{ \pm1
\right\}$ with eigenprojectors $P_\pm$. Thus \Ineqref{ineq:le2} becomes 
\begin{equation}
  \label{ineq:X3}
2P_++ i R\geq0,
\end{equation}
which shows that the overlap $\tr(P_-R)=-i\tr(\gamma_pR)$ must
vanish. As detailed in \cite{SuppInfo}, the matrix $R$ is the
  pointwise (Hadamard-) product of $\gamma_p$ with a
  symmetric matrix $r$: 
  $R_{kl}=r_{kl}(\gamma_p)_{kl}$, and the $r_{kl}$ are strictly negative
  whenever $d_k\not= d_l$.  Using the
  symmetry of $r$ and antisymmetry and purity of $\gamma_p$ then shows
  (see \cite{SuppInfo}) that this imposes $(\gamma_p)_{kl}=0$ whenever
  $d_k\not= d_l$, i.e.,  the direct-sum decomposition of $D$ and
  $\gamma_p$ into blocks of even dimension corresponding to 
constant $d_k$. \qed

A final simplification used in \Thref{th:th1} is that the pure state
of the environment can be taken to be the
vacuum state (with CM  $\gamma_E=J$). This is the case since for
$A=\sqrt{1-p}\id$ the FGCs $\cT_{A;p\gamma_E}$ and $\cT_{A;pJ}$ differ
only by unitary pre- and post-processing.

In summary, we have shown that there is 
only one family of degradable fermionic Gaussian channels, namely the attenuation channel $\cT_p:\gamma\mapsto(1-p)\gamma+pJ_{2n}$
(with losses $p\in[0,1/2]$). Hence FGCs have a much simpler
degradability structure than their qubit or  bosonic
Gaussian counterparts \cite{CRS08,CEGH08}.  In contrast to the case of
qubits, there are no degradable $n$-mode FGCs with large 
environment nor any (non-trivial)
Hadamard channels \cite[p.~196f]{Wilde2017}. Note that even channels
  very close to the ideal one such as $T_{A;B}$ with
  $A=\mr{diag}(\sqrt{1-x^2},\sqrt{1-y^2}), B=xy J$ or $A=\alpha\id,
  B=0$ are not degradable (unless $x=y$ or $\alpha=1$, respectively).

We can exploit the degradability of $\cT_p$ to compute its quantum
capacity, $Q(\cT_p)$, given by the channel's coherent information
\cite{Wilde2017}:
$Q(\cT_p)=\mr{max}_{\gamma}\left\{S(\cT_p(\gamma)-S((\cT_p\oplus\id)(\Gamma))\right\}$,
where $\Gamma$ is a purification of $\gamma$. That we can restrict to
Gaussian input is a consequence of the extremality of Gaussian states
as shown in \cite{WGC06,WPG07} (for bosons) and generalized to
fermions in \cite{Greplova2013} (see also \cite{SuppInfo}).
With $\gamma= \lambda J$ (general one-mode CM)
$\cT_p(\gamma)$ has eigenvalues $\pm i[p+(1-p)\lambda]$; we can take
$\Gamma =
\left( \begin{array}{cc} \lambda J & \sqrt{1-\lambda^2}X\\
    -\sqrt{1-\lambda^2}X &\lambda J
\end{array} \right)$ and find that $(\cT\oplus\id)(\Gamma)$ has eigenvalues $\pm
i$ (one pure mode) and $\pm
i(1-p+p\lambda)$. 
This reduces the computation of $Q$ to a simple one-parameter
optimization: 
$Q(p) = \max_{-1\leq\lambda\leq1}\left\{H(\frac{(1-p)(1-\lambda)}{2})-H(\frac{p (1-\lambda)}{2})\right\}$, where $H(p)=-p\log p-(1-p)\log(1-p)$ is the binary
entropy.
The channel $\cT_p$ is equivalent to the qubit amplitude damping channel with Kraus operators
$K_1=\proj{0}+\sqrt{(1-p)}\proj{1}$ and $K_2=\sqrt{p}\ketbra{0}{1}$ whose
quantum capacity was computed in \cite{GiFa05}. Notably, the \emph{classical} capacity \cite{Wilde2017} of the qubit amplitude
damping channel remains unknown to-date, although lower \cite{GiFa05} and upper
bounds \cite{wang2018semidefinite} have been obtained.

There are several interesting directions for further research: 
(1) The generalization of the above result to the case of $n\to m$
channels is important, in particular for the
\emph{antidegradability} even of $n\to n$ channels, since, in
general, the complementary channel is a map between systems of
different number of modes.  While it is clear that $n\to m$ channels
with $m<n$ are never 
degradable (since $A$ has a kernel), in the case $m>n$, the positivity
of the real part of \Eqref{eq:poscond4} is no longer 
easy to decide
since it may depend on the details of $\gamma_E$ and examples of
degradable channels with Choi rank larger than $\max\{n,m\}$ exist 
(see \cite{SuppInfo}). Moreover, the uniqueness of the degrading map is no
longer ensured and all such maps (both Gaussian and, possibly, non-Gaussian)
would need to be examined. (2) FGCs may provide a simple
  setting to search for exclusively conjugate-degradable channels
  \cite{Bradler2010} and (3) our result on degradability of FGCs may
  be of use in  bounding the private and quantum capacity  of some
  non-degradable channels or non-Gaussian channels by 
exploiting the notion of  approximate degradability \cite{SSR14} or
following \cite{WPG07} via a fermionic Gaussian teleportation channel.

\begin{acknowledgments}
We thank an anonymous referee for drawing our attention to the matter
of uniqueness of the degrading map. \\
  We acknowledge the Centro de Ciencias de Benasque Pedro
  Pascual, where the central part of this work was completed.  GG
  thanks the Institute for Mathematical Sciences, National University
  of Singapore for the invitation to the 2013 \emph{Mathematical
    Horizons of Quantum Physics} workshop, during which early parts of
  this work were done. EG acknowledges financial support of the Villum
  Foundation and the Bakala Foundation.
\end{acknowledgments}

%

\end{bibunit}

\pagebreak

\begin{bibunit}[apsrev4-1]
\newcommand{\tPhi}{\tilde{\Phi}}
\newcommand{\tPsi}{\tilde{\Psi}}

\begin{widetext}
\begin{center}
\textbf{\large Supplementary material: Degradability of Fermionic Gaussian Channel}
\end{center}
\end{widetext}

\setcounter{equation}{0}
\setcounter{figure}{0}
\setcounter{table}{0}
\setcounter{page}{1}
\setcounter{lemma}{0}
\setcounter{definition}{0}
\setcounter{theorem}{0}
\makeatletter
\renewcommand{\thepage}{S\arabic{page}}
\renewcommand{\theequation}{S\arabic{equation}}
\renewcommand{\thefigure}{S\arabic{figure}}
\renewcommand{\thelemma}{S\arabic{lemma}}
\renewcommand{\thetheorem}{S\arabic{theorem}}
\renewcommand{\thedefinition}{S\arabic{definition}}
\renewcommand{\theexample}{S\arabic{example}}
\renewcommand{\thecorollary}{S\arabic{theorem}.\arabic{corollary}}
\renewcommand{\bibnumfmt}[1]{[S#1]}
\renewcommand{\citenumfont}[1]{S#1}

In this Supplementary Material we collect some simple statements and
details of the proofs used in the main text. After some useful general
lemmas on the properties of block matrices we provide technical
details for the proof of the degradability theorem. We conclude with
stating and proving an extremality theorem for fermionic Gaussian
states. All equations and lemmas of the Supplementary Material are labeled with an
``S'' in front, other labels refer to the main text.

\section{Some useful lemmas on matrices}
\begin{lemma}[Positivity conditions]\label{Sle:Schur}
(i)  For the Hermitian matrix 
\[W=\left( \begin{array}{cc} X&Y\\ Y^\dag&Z,
\end{array} \right)\] 
defined on (for our purposes finite dimensional) $\cH=\cH_A\oplus\cH_B$, we have that $W\geq0$ if and only if 
\begin{equation}\label{Seq:Schur}
\mr{ker}(Z) \subseteq \mr{ker}(Y) \,\,\mbox{ and  }\,\, 
X - YZ^{-1}Y^\dag\geq0. 
\end{equation}
(ii) For the case of \emph{real} matrices $X,Y,Z$ with $Z=X$ and $Y=-Y^T$ 
we also have that $W\geq0$ if and only if $X+iY\geq 0$.
\end{lemma}
\Proof: (see \cite{GKLC01b})

By definition $W\geq0$ iff ${a \choose b}^\dag W {a \choose
  b}\geq0\ \forall a\in\cH_A, \forall b\in\cH_B$, i.e., iff
\begin{equation}
  \label{eq:leS1}
  a^\dag Xa+b^\dag  Zb+a^\dag Yb+b^\dag Y^\dag a\geq0,
\end{equation}
from which the necessity of the kernel-condition in  \Eqref{Seq:Schur}
is evident by taking $b\in\ker Z$ but $\not\in\ker Y$. 
Furthermore, if $\mr{ker}(Z) \subseteq \mr{ker}(Y)$ then
$Z^{-1}Y^\dag$ is well-defined and inserting $b=-Z^{-1}Y^\dag a$ in
\Eqref{eq:leS1} yields the inequality in \Eqref{Seq:Schur}.\\
Sufficiency follows since the kernel-condition of \Eqref{Seq:Schur}
allows us to define $\tilde{a}=Z^{-1}Y^\dag a$ and deduce
$Z\tilde{a}=Y^\dag a$; rewriting \Eqref{eq:leS1} using $\tilde{a}$ we
obtain that its left-hand side is $a^\dag \left(X -
  YZ^{-1}Y^\dag\right)a + (b+\tilde{a})^\dag Z (b+\tilde{a})$, both of
which are positive by \Eqref{Seq:Schur}. \\
The second statement follows by observing that $W$ is real and thus
positive if $(a\oplus b)^\dag W(a\oplus b)\geq 0$ for all real vectors
$a,b$ and that $(a+ib)^\dag (X+iY)(a+ib)=(a\oplus b)^\dag W(a\oplus
b)$. \qed

This is useful for confirming whether an antisymmetric matrix $M$
  is a valid CM as follows: Consider
  \begin{equation}
    \label{eq:CJ-CM}
    M\equiv M_{(A;B)} = \left( \begin{array}{cc} B& A\\-A^T& 0
\end{array} \right).
  \end{equation}
$M$ is a CM iff $\id-MM^T\geq0$ which is
  by (i) the case iff $\left( \begin{array}{cc} \id & M\\-M^T&\id
\end{array} \right)\geq0$, which, in turn, holds by (ii) iff
$\id+iM\geq0$. This matrix has the form
\begin{equation}
\id+iM=\left( \begin{array}{cc} \id + iB & iA\\-iA^T&\id
\end{array} \right)
\end{equation}
and we can see that it is Hermitian. Applying Schur complement (i) again we see that $M$ is valid covariance matrix iff  $\id-AA^T+iB\geq 0$.

\begin{lemma}[Bipartite Gaussian Unitaries]\label{Sle:BipartiteUnitaries}
  Let $U$ be a Gaussian unitary acting on $n+m$ modes and let the
  orthogonal matrix 
\[
O = \left( \begin{array}{cc} O_{11}&O_{12}\\ O_{21}&O_{22}
\end{array} \right)
\]
be its phase space representation. Then we can find orthogonal
operations $Q_i,R_i, i=1(2)$ acting on the first (second) set of modes
such that (we assume $n\leq m \equiv n+k$)

\begin{equation}
\label{eq:suppO}
O = (Q_1\oplus Q_2) \left(  \begin{array}{c|cc} 
D & \sqrt{\id - D^2} & 0\\
\hline
-\sqrt{\id-D^2} & D & 0\\
0 & 0 & \id_{2k}
\end{array} \right) (R_1\oplus R_2),
\end{equation}
where $\id\geq D\geq0$ is a diagonal $n\times n$ matrix.
 \end{lemma}
\Proof: Equation \eqref{eq:suppO} is an immediate consequence of the orthogonality conditions and the
singular value decomposition 
\cite{HJ87}  of the real matrices $O_{ij}$. In particular, for
singular value decomposition of $O_{11}=K_1D_1L_1$ we find
$Q_1=K_1^{-1}$ and $R_1=L_1^{-1}$ and so on for all the remaining
blocks. In the other words, we pick the local orthogonal
transformations such that they "undo" the singular value
decomposition. On the simplified matrix we impose the requirements
given by the fact that $O$ has to belong to the special orthogonal
group ($OO^{\dagger}=\id$ and $O^{\dagger}O=\id$ and $\det O=1$). These directly provide the reduction given by \eqref{eq:suppO}.\qed

If $n<m$ this shows that the $n$ modes of the first system interact
with no more than $n$ modes of the second system (those in the range
of $O_{12}$) while the modes in the kernel of $O_{12}$ are ``spectator
modes''.  Their only influence on the state in the first system after
application of $O$ stems from their prior entanglement with the
``interacting'' modes.

\section{Existence, Uniqueness and Gaussianity of the Degrading Map}

Given a quantum channel $\cT$, the complementary channel $\cT^c$ is
not unique \cite{Bra15}. Since (anti)degradability is defined via the existence of
\emph{a} CPM connecting $\cT$ and $\cT^c$,  it is necessary to exclude
the possibility of \emph{any} of these maps being completely positive
in order to prove non-(anti)degradability. In particular, this
requires to consider both the possibility of Gaussian and non-Gaussian
degrading maps. The following lemmas show that for FGCs we can
concentrate on Gaussian degrading maps.

\begin{lemma}[Degradability independent of choice of complementary
  channel]\label{Sle:unique} 
  Let $\cT:\cB(\cH_A)\to\cB(\cH_B)$ be a quantum channel between
  finite dimensional Hilbert spaces and $\cT^c$ a
  channel complementary to it. Then $\cT$ is degradable iff there
  exists a trace-preserving (TP) completely positive map (CPM)
  degrading $\cT$ to $\cT^c$.
\end{lemma}
\Proof: 
We use the Stinespring representation of $\cT$ \cite{Wolf2012}, which
is unique up to partial isometries and gives 
rise to a set of complementary channels. Let $r_{\cT}$ be the Kraus rank of $\cT$, i.e., the minimal number of Kraus operators needed to represent
$\cT$. Then 
for all $r\geq r_{\cT}$ there is a partial isometry $V:\cH_A\to\cH_B\otimes\cH_{E'}$
with $\dim\cH_{E'}=r$ such that 
\[
\cT(\rho)= \tr_{E'}\left( V \rho V^\dag\right)
\]
and
$T^{c\prime}(\rho)=\tr_B\left( V \rho V^\dag\right)$ defines a
complementary channel. It is related to the \emph{minimal} complementary
channel by a partial isometry $W:\cH_{E,\min}\to\cH_{E'}$. 
We show that there is a TPCP degrading map $\cD'$ with $\cD'\circ\cT =
\cT^{c\prime}$ iff there is one degrading $\cT$ to the minimal
complementary channel $\cT^c$.\\ 
``If'' is clear, since the isometry $W$ defines a TPCPM
$\Phi_W:\rho\mapsto W\rho W^\dag$ and if there is a 
quantum channel $\cD$ degrading $\cT$ to $\cT^c$ then
$\cD'=\Phi_W\circ\cD$ is a quantum channel and degrades to
$\cT^{c\prime}$. To see ``only if'', we construct a TPCPM 
$\Phi_{W,\mr{inv}}$ such that 
\[\cT^c = \Phi_{W,\mr{inv}}\circ\cT^{c\prime}. \]
Given $\Phi_{W,\mr{inv}}$ a quantum channel $\cD'$ degrading $\cT$ to
$\cT^{c\prime}$ yields also the TPCPM
$\cD=\Phi_{W,\mr{inv}}\circ\cD'$ which degrades $\cT$ to $\cT^c$.\\
To construct $\Phi_{W,\mr{inv}}$ we use that while $W$ is not
invertible, there is a one-to-one relation between vectors in $\cH_E$
and the subspace $\cH_W=W\cH_E\subset\cH_{E'}$. Denote by
$P_W=WW^\dag$  the orthogonal projector on $\cH_W$. 
Using $W^\dag$ we define 
\[
\Phi_{W,\mr{inv}}(x) =  W^\dag P_{W}x P_{W}  W + \tr(x P_{W}^\perp) \proj{\psi_0},
\]
where $\ket{\psi_0}\in\cH_{E}$ is a pure state. It is straightforward
to check that $\Phi_{W,\mr{inv}}$ is TP and CP: its Kraus operators
are $K_0=W^\dag P_W$ and $K_k=\ketbra{\psi_0}{k}$ for an
orthonormal basis $\left\{ \ket{k}:k\geq 1 \right\}$ spanning the
orthogonal complement of $\cH_W$ and $\sum_k K_k^\dag K_k =
P_W+P_W^\perp=\id_{\cH_{E'}}$. \qed

\Leref{Sle:unique} allows to study degradability for a fixed choice of
dilation and to conclude from a proof that if there is no degrading map
for a given dilation (and corresponding complementary channel) than
the channel is not degradable. However, we still have to deal with the
non-uniqueness of the degrading map for fixed complementary
channel. From now on, we concentrate on fermionic Gaussian channels
$\cT \equiv \cT_{(A;B)}$.

First, we note that if the matrix $A$ has a kernel, then no degrading
map exists, since we can readily find two distinct Gaussian input
states with CM $\gamma1\not=\gamma_2$ such that
$\cT_{(A;B)}(\gamma_1)=\cT_{(A;B)}(\gamma_2)$, while
$\cT^c_{(A;B)}(\gamma_1)\not=\cT^c_{(A;B)}(\gamma_2)$. Assuming
(without loss of generality, see \Leref{Sle:UnitConcat}) that $A=0\oplus
A'$ the CMs $\pm J_2 \oplus\gamma'$ are both mapped to the output
state $0_2\oplus A' \gamma' (A')^T+B$, but to different complementary
outputs. In particular, this rules out the existence of $n\to m<n$
degradable FGCs.

Now it remains to show that in those cases, where a Gaussian degrading
map can be constructed ($n\to n$; ker$A=\emptyset$) it is unique (up
the CP maps) so that we can conclude from it being not completely positive
that no such degrading map can be found.

\begin{lemma}[Uniqueness and Gaussianity of the degrading map for $n\to n$ FGCs with
  ker$A=\emptyset$]\label{Sle:uniqueDegMap-FGC}  
The degrading map for $n\to n$ FGCs $\cT_{(A;B)}$ with invertible $A$ is unique (up to partial
isometries as discussed in \Leref{Sle:unique}) and is given by the
Gaussian map constructed in Eqs.~(11,12) 
of the  main text.
\end{lemma}

\Proof: As shown in \cite{Bra15}, the degrading map for a
  $\cB(\cH)\to\cB(\cH')$ channel is unique if (i) the superoperator
  representing the channel has maximal rank and (ii)
  $\dim\cH\geq\dim\cH'$.\\
  We consider $n\to n$ channels, so (ii) always holds. To show (i), we
  use the special properties of Gaussian channels $(A,B)$: whenever
  $A$ is invertible, we can immediately construct the inverse of the
  channel as the Gaussian fermionic map $(A^{-1},-A^{-1}BA^{-T})$.
  For Gaussian states it is immediate to see that this map inverts the
  channel. To show that this implies that it inverts it for \emph{all}
  input states, we prove in the following \Leref{Sle:gaussianbasis}
  that the Gaussian states span the space of all density
  matrices. Therefore, by linearity any map is fixed by its action on
  Gaussian states and it suffices to construct the inverse on those
  states. Clearly, the superoperator that implements an invertible map
  is invertible and thus has full rank. \qed

\begin{lemma}[Gaussian states span the space of density matrices]
\label{Sle:gaussianbasis}
  Every density matrix corresponding to a fermionic state (FS) of $n$ modes
  can be expressed as a linear combination of density matrices of
  Gaussian fermionic states.
\end{lemma}

\Proof: Note that FS are mixtures of even and odd states \cite{Brav05}. Since the
latter can be transformed into the former by a Gaussian operation (or
by adjoining a single occupied mode), it suffices to show the Lemma
for the even states. These are spanned by 
$\ketbra{m}{m'}\pm\ketbra{m'}{m}$ with $m,m'$ bitstrings that differ
at an even number of locations. For $n=2$ the projectors on the
(unnormalized) Gaussian states $\ket{00},\ket{11},
\ket{\Phi_\pm}=\ket{00}\pm\ket{11},|\tilde{\Phi}_{\pm}\rangle=\ket{00}\pm
i\ket{11}$ span the set of FS, as is easily confirmed by obtaining the
(only possible) off-diagonal elements
$\ketbra{11}{00}\pm\ketbra{00}{11}$ as the superpositions
$\proj{\Phi_+}-\proj{\Phi_-}$ and
$\proj{\tilde{\Phi}_+}-\proj{\tilde{\Phi}_-}$. Now we proceed by
induction. Assume the Lemma holds for $2n$-mode states. Then for
$2n+1$ modes no new correlations appear since any operator 
$\ketbra{m}{m'}\pm\ketbra{m'}{m}$ on an odd number of modes has (due
to $m,m'$ having the same parity) at least one mode that factorizes
($m_j=m'_j$) and thus we can use Gaussian unitaries (GU), namely swaps
and particle-hole flips, to map $m,m'$ to $r,r'$ such that the last
mode factorizes (and is, without loss of 
generality) empty. Thus the matrix
element under consideration is obtained by combining projectors on
$2n+1$-mode Gaussian states of the form $\ket{G_{2n}}\ket{0}$, where
$\ket{G_{2n}}$ are two mode Gaussian states.  \\
To conclude the induction we need to consider the case of $2n+2$ modes
in which new correlations can appear. Again we can use GU to map
$\ketbra{m}{m'}$ to either $\ketbra{r00}{r'00}$ (expressions that can
be obtained from states in which the last two modes factorize) or
$\ketbra{0\dots0}{1\dots1}$.  The former lie in the span of the
projectors on Gaussian states $\ket{G_{2n}00}$ (and GU-transforms
thereof) and the latter is obtained by adding the (projectors on
  the) Gaussian FS
$\ket{\Phi_{\pm}}_{12}\ket{\Phi_{\pm}}_{34}\dots\ket{\Phi_{\pm}}_{2n+1,2n+2}$
and
$|\tPhi_{\pm}\rangle_{12}|\tPhi_{\pm}\rangle_{34}\dots|\tPhi_{\pm}\rangle_{2n+1,2n+2}$. \qed

\section{Standard Form and Complete Positivity of the Gaussian
  Degrading Map}

\begin{lemma}[Degradability unaffected by concatenation with unitaries] \label{Sle:UnitConcat}
  The quantum channel $\cT$ is (anti)degradable iff for any two
  unitary channels $\cU_1,\cU_2$ the
  channel $\cT'= \cU_2\circ \cT\circ \cU_1$ is (anti)degradable.
\end{lemma}
\Proof: If $\cD$ degrades $\cT$ to one of its complementary channels $\cT_c$
then $\cD'= \cD\circ\cU_2^{-1}$ degrades $\cT'$ to its complementary
channel  $\cT'_c=\cT_c\circ\cU_1$. Since $\cU_i$ are invertible the
converse also holds.\qed

\begin{lemma}[Degradability of Composite Channels]\label{Sle:DegCompChan}
  Two quantum channels $\cR,\cS$ are both degradable if and only if
  the combined channel 
  $\cR\otimes \cS$ is degradable. 
\end{lemma}

\Proof: If $\tilde{\cR},\tilde{\cS}$ are degrading TPCPMs for $\cR,\cS$, resp., then
$\tilde{\cT}=\tilde{\cR}\otimes\tilde{\cS}$ is also TPCP and is the degrading map for
$\cT=\cR\otimes \cS$. 
For the converse case, we can without loss of generality consider a unitary
  dilation $U_{RE_R}\otimes U_{SE_s}$ of the product channel
  $\cT=\cR\otimes\cS$ since the degradability does not depend on how we use
  the isometric freedom in the choice of the complementary channel
  (see \Leref{Sle:unique}). With this
  choice, $\cT^c=\cR^c\otimes \cS^c$ and if $\tilde{\cT}$ is a CPM degrading $\cT=\cR\otimes \cS$
then  
the CPMs $\tilde{\cR}=\cR_2\circ\tilde{\cT}\circ \cR_1$ (with
$\cR_1:\rho\mapsto\rho\otimes \cS(\Phi)$ and
$\cR_2:\rho_{1'2'}\mapsto\tr_{2'}(\rho_{1'2'})$ for an arbitrary state
$\Phi$) and $\tilde{\cS}$ (defined
similarly) satisfy $\tilde{\cR}\circ
\cR(\rho)=\tr_{2'}\left( \tilde{\cT}(\cR(\rho)\otimes \cS(\Phi))
\right)=\tr_{2'}\left( \tilde{\cT}(\cT(\rho\otimes \Phi)) \right)
=\tr_{2'}\left( \cT^c(\rho\otimes \Phi) \right)=\cR_c(\rho)$ (and
analogously for $\tilde{\cS}$), i.e., they degrade $\cR$ resp. $\cS$.\qed

We would like to apply this theorem to composite (Gaussian) fermionic
  channels, i.e., channels for which the $A$ and $B$ matrices
  characterizing the phase-space transformation have direct-sum
  character $A=A_1\oplus A_2; B=B_1\oplus B_2$. Then the above theorem
  applies as well since fermionic channels applied to different modes
  commute with each other \cite{SSGK18}, e.g., 
  $\tilde{\cT}\circ\cT=\tilde{\cR}\otimes\id\circ\cR\otimes\id\circ\id\otimes\tilde{\cS}\circ\id\otimes\cR=\cR^c\otimes\cS^c$
  and $\cT_{(A;B)}=\cT_{(A_1;B_1)}\otimes \cT_{(A_2;B_2)}$ etc. Here
  ``$\otimes$'' is used to indicate on which of the two independent
  sets of modes these channels act.

\begin{lemma}[Perfectly transmitted Majorana modes]\label{Sle:perfectmode}
A FGC with $k\geq2l$ perfectly transmitted Majorana modes is
unitarily equivalent to a FGC of the form $T=T_{(\id_{2l};0)}\oplus
T_{D',B'}$.   
\end{lemma}

  \Proof: Consider a FGC $T_{A;B}$ in standard form. It transmits $k$
  Majorana modes ``perfectly'', if the (diagonal) matrix $A$ has at
  least $k$ singular values 1. In this case, the CPM-condition
  [Eq.~(4) of the main text]
  implies that $B$ must vanish on all those modes. Let $l$ be the
    largest integer such that $k\geq 2l$. Then the
  channel is a particular case of a Gaussian product channel $(A; B) =
  (A_1\oplus A_2 ; B_1\oplus B_2)$ with $A_1 = \id_{2l}$ and $B_1=0$
and where $A_2$ now has at most one singular value equal to 1.\qed 
\vspace{3mm}

This will be useful for the degradability classification later, since by
\Leref{Sle:DegCompChan} it then suffices to study the smaller channel $T_{D',B'}$.

An interesting characteristic of the channel is the \emph{smallest}
environment such that a physical representation with a pure state is
possible. The dimension of that environment is called the \emph{Choi
  rank} of the channel \cite{CRS08} and is given by the rank of the
Choi-Jamiolkowski state of the channel. For our fermionic case we
consider the state obtained by letting the channel $T_{A;B}$ act on
the first $n$ modes of a $2n$-mode maximally entangled fermionic
Gaussian (e.g., with CM as in Eq.~(2) in the main text 
with $l=0$ and $\lambda_j=1$).  In the following, it is convenient to
measure the Choi rank via the \emph{minimum number of fermionic modes}
needed to purify this Gaussian state.  For an $n\to n$ FGC $T_{A;B}$
we show in the following lemma that its Choi rank is
$2n-\mr{dim}\ker(\id-AA^T-iB)$:

\begin{lemma}[Choi rank of $n\to n$ Gaussian channels]\label{Sle:Choirank}
For an $n\to n$ channel $T_{(A;B)}$, its Choi rank is 
given by half the number of eigenvalues of $M_{(A;B)}$ with modulus smaller
than 1, that is, by $2n-l$ modes where 
$l=\mr{dim}\ker(\id-iM_{(A;B)})=\mr{dim}\ker(\id-AA^T-iB)$. If
$\id-AA^T$ has full rank then $l=\mr{dim}\ker(\id-iB')$. 
\end{lemma}

\Proof: 
The CJ-state of the channel $T_{A;B}$ is the $2n$-mode Gaussian state
$\rho_{M_{A;B}}$ and since $M=O(\oplus_k \lambda_k J)O^T$, it is
unitarily equivalent to $\rho_{\oplus_k \lambda_k J}$, which is pure
for the $l$ modes with $|\lambda_k|=1$ and a full-rank (thermal) state
on the rest. Since $\dim\ker(\id-i\lambda J)$ is 1 if $|\lambda|=1$ and
$2$ otherwise, we have $l=\dim\ker(\id-iM_{(A;B)})$. The expression in
terms of $A$ and $B$ follows since $(\id-iM_{(A;B)}){x \choose y}=0$ iff
$y=-iA^Tx$ and $x\in\ker(\id-iB-AA^T)$. \\
If $\id-AA^T$ has full rank, then $W=(\id-AA^T)^{-1/2}$ is a well-defined
similarity transformation and hence the kernels of $(\id-AA^T-iB)$ and
$W(\id-AA^T-iB)W^T=(\id-iB')$ have the same dimension $l$. \qed

  \vspace{3mm}

The following three Lemmas constitute the core of the proof of our main
theorem. 

\begin{lemma}[Small Choi-rank necessary for degradability]
  \label{Sle:ChoiDeg}
  A $n\to n$ FGC is not degradable unless its Choi rank
  is $\leq n$ modes. 
\end{lemma}
\Proof:  The positivity of $\tilde{M}=\id-\tilde{A}\tilde{A}^T-i\tilde{B}$
[cf. Eqs.~(11-13) of the main text] can be 
checked using \Leref{Sle:Schur}. 
To apply it, we view $\tilde{M}$ as a $2\times2$ block matrix
with blocks $\tilde{X},\tilde{Y},\tilde{Z}$ and also write the CM of the pure
environmental state (which appears in $\tilde{B}$) in that form:
\[\gamma_E=\left( \begin{array}{cc} E_{11}&E_{12}\\
    -E_{12}^T&E_{22} \end{array} \right).\] 
Then we have 
\begin{align*}
  \tilde{X}&=2\id-D_A^{-2}-iD_AE_{11}D_A + i(D_A^{-1}+D_A)E_{11}(D_A^{-1}+D_A),\\
\tilde{Y}&=-iD_AE_{12},\\
\tilde{Z}&=\id-iE_{22}.
\end{align*}
Now we consider a
minimal unitary dilation of $T_{A;B}$ on $2n-l$ environmental modes. We use
for the pure $n\times(n-l)$ bipartite FGS $\gamma_E$  
the standard form as given in Eq.~(2) in the main text and write
\begin{align*}
 E_{11} &= Q_1(J_{2l}\oplus L) Q_1^T,\\
 E_{22} &= Q_2LQ_2^T,\\
 E_{12} &= Q_1\left(\begin{array}{c}0_{2l\times 2(n-l)}\\ K\end{array}\right)Q_2^T,
\end{align*} 
with $Q_1\in \SO(2n), Q_2\in \SO(2n-2l)$ and
$L=\oplus_{j=1}^{n-l} \lambda_j J$, $K = \oplus_j \kappa^{n-l}_j X$ with
$\lambda_j^2+\kappa_j^2=1$. The blocks refer to the
$l\leq n$ pure and $n-l$ mixed modes in $B'$ (i.e., those environmental modes
coupled to 
directly by the system) and the $n-l$ environmental spectator modes
needed to purify $B'$. Hence we have $|\lambda_j|<1, |\kappa_j|>0$.
Therefore, $\tilde{Z}=\id-iE_{22}=Q_2\left(
  \id-iL \right)Q_2^T$ has no kernel, i.e., we
can use the Schur complement formula \Eqref{Seq:Schur} to check positivity of
$\tilde{M}$. The term $\tilde{Y}\tilde{Z}^{-1}\tilde{Y}^\dag$ simplifies to
$D_AQ_1\left( \begin{array}{cc} 0_{2l} & \\ &  K (\id-iL)^{-1}K^T
  \end{array} \right)Q_1^TD_A$ and using the block-diagonal form of $K,L$
we find $K(\id-iL)^{-1}K^T=\id-iL$, giving the condition
\begin{equation}
  \label{Seq:poscond3}
  2\id-D_A^{-2}-D_AQ_1\left( \begin{array}{cc} 0&\\ & \id_{2(n-l)}
\end{array} \right)Q_1^TD_A+i(\dots)\geq0,
\end{equation}
where the imaginary part is not relevant in the following. 
If $l=0$ (no pure modes), the condition for the positivity of
$\tilde{M}$ simplifies to $2\id-D_A^{-2}-D_A^2+i(\dots)\geq0$,
and the real part has negative eigenvalues (unless
$D_A=\id$, a case we 
excluded by \Leref{Sle:DegCompChan}), thus \Ineqref{Seq:poscond3} does not
hold and the channel is not degradable in that case.  To extend this
to all $l<n$ we  
multiply \Ineqref{Seq:poscond3} from both sides with $D_A^{-1}$ (we
already know $D_A\geq \id/\sqrt{2}$) and the real part
of the LHS becomes
\begin{equation}
  \label{Seq:poscond4}
  2D_A^{-2}-D_A^{-4}-Q_1(0_{2l}\oplus\id_{2(n-l)})Q_1^T.
\end{equation}
To see that this cannot be positive unless $l=n$ (pure environment) 
we use a simple condition on the eigenvalues of two
Hermitian matrices and their sum implied by Horn's conjecture,
\cite{Horn62,KnTa01}: Let $\lambda_i,\mu_j,\nu_k$ denote the descendingly
ordered eigenvalues of the Hermitian matrices $X,Y,X+Y$, respectively.  Then
we have \cite{KnTa01}
\begin{equation}
  \label{Seq:Horn1}
  \nu_k\leq\lambda_i+\mu_j\,\,\,\forall i+j=k+1.
\end{equation}
For our purposes we take $X=2D_A^{-2}-D_A^{-4}$ and
$Y=-Q_1(0_{2l}\oplus\id)Q_1^T$ and pick
$j=2l+1$ such that $\mu_j=-1$. Then for all $i>1$ we have
\[
  \nu_{2l+i}\leq\lambda_i+\mu_{2l+1}=\frac{2}{d_i^2}-\frac{1}{d_i^4}-1\,\,\,\forall i=1,\dots,2(n-l).
\]
Unless $n-l=0$ (pure environment) we can take $i=2$ which means that $d_i<1$ in
which case the expression on the RHS is negative.\qed

\begin{lemma}[Only constant-loss channels are degradable] \label{Sle:ConstLoss}
  A $n\to n$ FGC $T_{(D;B)}$ in standard form with
  Choi rank $n$ is degradable if and only if $D=\oplus_j (d_j\id_{2n_j})$
  (i.e., all eigenvalues $d_j$ of $D$ have an even degeneracy), $d_j\geq1/\sqrt{2}$, and
  $B=\oplus_j B_j$ ($B$ does not contain correlations between blocks
  pertaining to nondegenerate values: $d_j\not= d_k\implies
  B_{jk}=0$). 
\end{lemma}
\Proof: 
We have $A=D$ diagonal and
$B=\sqrt{\id-D^{-2}}\gamma_p \sqrt{\id-D^{-2}}$ for a $n$-mode pure-state CM
$\gamma_p$; the degrading map is given by $\tilde{A}=\sqrt{\id-D^{-2}}$ and
$\tilde{B}=D\gamma_p D-(D^{-1}-D)\gamma_p(D^{-1}-D)$ and the condition for
degradability becomes 
\begin{equation}\label{Seq:X}
2\id - D^{-2}+i\left[  D\gamma_p D-\left(\frac{\id}{D}-D\right)\gamma_p\left(\frac{\id}{D}-D\right)\right]\geq0.
\end{equation}
We show now that this is only the case if $D\geq\id/\sqrt{2}$ and if
$(\gamma_p)_{kl}=0$ whenever $d_k\not= d_l$, i.e., 
$\gamma_p$ is a direct sum of independent pure states $\oplus
\gamma_{p,k}$ and $D$ a corresponding direct sum of terms proportional
to $\id$.  We already saw that $D\geq\id/\sqrt{2}$ is necessary for
degradability. 

(a) If there is an eigenvalue equal to $1/\sqrt{2}$, say
$d_i=1/\sqrt{2}$, then the
real part of \Ineqref{Seq:X}  has a kernel and therefore all the $ij$-th entries of the imaginary part must vanish.
The $ij$-th entry of the term in square brackets is
$d_id_j[1-(d_i^{-2})(1-d_j^{-2})]\gamma_{ij}$ and putting
$d_i=1/\sqrt{2}$, we see immediately that the inequality can only
hold if $\gamma_{ij}=0$ for all $j$ such that $d_j\not=d_i=1/\sqrt{2}$. I.e., a
channel with some $d_i=1/\sqrt{2}$ can only be degradable if $D=D'\oplus
\frac{1}{\sqrt{2}}\id$ and $\gamma=\gamma_1\oplus\gamma_2$ in accordance with
our theorem.

(b) Then, assuming $D>\id/\sqrt{2}$ and multiplying \Eqref{Seq:X} by $\frac{1}{\sqrt{2\id-D^{-2}}}$ from left and
right and introducing $W=\sqrt{2D^2-\id}$, we see that the inequality is equivalent 
to
\begin{equation}\label{Sineq:X2}
\id+ i \Inv{W}\left[\gamma_p-
 D^2\gamma_p-\gamma_pD^2 \right]\Inv{W}\geq0.
\end{equation}
(c) To see that this necessitates the block structure of $\gamma_p$,
we write the imaginary part of \Eqref{Sineq:X2} as
$\gamma_p+R$ and exploit that $\gamma_p$ is pure, which
implies that $i\gamma_p=P_+-P_-$ with orthogonal projectors $P_\pm$ on
the $\pm1$-eigenspaces of $\gamma_p$ that span the full space
($P_++P_-=\id$). Thus,   
\Eqref{Sineq:X2} is equivalent to 
\begin{equation}\label{Sineq:X3}
2P_++ i R\geq0,
\end{equation}
which implies that $\tr(P_-R)=0$. Since the antisymmetric matrix
$R$ is also traceless ($\tr(R)=0$) we obtain that it is necessary for
\Eqref{Seq:X} that 
\begin{equation}
  \label{Seq:X4}
  \tr(\gamma_p R)=-i\tr[(P_+-P_-)R]=-i\tr[(P_++P_-)R]=-i\tr(R)=0.
\end{equation}
(d) This condition is sufficient to complete the proof. Let us consider
$\gamma$, $R$ as block matrices with $2\times2$ blocks denoted by $\gamma^{ij},
R^{ij}, i,j=1,\dots, n$ and with entries $\gamma^{ij}_{kl}, k,l=1,2$
etc. We then use that the elements of $R$ are multiples of the corresponding
element of $\gamma_p$:
\[
R^{ij}_{kl} =
\left(-1+\frac{1-(d^i_k)^2-(d^j_l)^2}{\sqrt{2(d^i_k)^2-1}\sqrt{2(d^j_l)^2-1}}\right)\gamma^{ij}_{kl}
\equiv r^{ij}_{kl}\gamma^{ij}_{kl}, 
\]
and it is easily checked that $r^{ij}_{kl}<0$ for $d^i_k\not= d^j_l>1/\sqrt{2}$. 
To see this we write $d^i_k=d+x$, $d^j_l=d-x$ (we have:
$d>1/\sqrt{2}$ and $(1-1/\sqrt{2})/2\geq x\geq0$) and note that
$r^{ij}_{kl} = -1+\frac{1-2(d^2+x^2)}{\sqrt{[1-2(d^2+x^2)]^2-16x^2d^2}}$. The
second term is  negative since $d^2+x^2>(d-x)^2>1/2$ and the
denominator is positive.
The matrices $\gamma$ and $R$ are antisymmetric, hence
$\gamma^{ji}=-(\gamma^{ij})^T$ and $r^{ji}_{kl}=r^{ij}_{lk}$,  which allows to
simplify the expression for $\tr(\gamma_p R)$:
\begin{align*}
  \tr(\gamma_p R)&=\sum_{i,j} \tr(\gamma^{ij}R^{ji})\\
&=\sum_i \tr(\gamma^{ii}R^{ii})+\sum_i\sum_{j\not=i}\tr(\gamma^{ij}R^{ji})\\
&=-2\sum_i r^{ii}_{12}(\gamma^{ii}_{12})^2-\sum_i\sum_{j\not=i}\sum_{k,l=1}^2r^{ji}_{kl}(\gamma^{ij}_{lk})^2.
\end{align*}
(e) With all $r^{ij}_{kl}\leq0$, all terms in the
above sums are 
$\geq0$ and the sum is strictly positive unless \emph{all} the
terms are zero which requires that
$\gamma^{ij}_{kl}=0$ whenever $d^i_k\not=d^j_l$. Since the $d^i_k$ are
monotonically decreasing in standard form, we have
$D=\oplus_m d_m \id_{n_m}$ and thus \Eqref{Seq:X4} implies $\gamma_p=\oplus_m
G_m$. The blocks $G_m$ are 
(of course) antisymmetric and must all have even dimension: odd blocks
are incompatible with $\gamma_p$ being pure, since the eigenvalues of
$i\gamma_p$ are $\{\pm1\}$ and must coincide 
with those of the blocks $iG_m$ -- but an odd antisymmetric matrix
always has at least one eigenvalue zero. Thus all the $G_m$ are
valid CMs and we have the advertised block structure of $\gamma_p$ and $D$.\qed

\begin{example}[A large-Choi-rank $n\to m$ degradable FGC]
If $m>n$ then in standard form 
$A=\left( \begin{array}{cc} D & 0\end{array} \right)^T$ (for a
$2n\times 2n$ matrix $D$) and 
\Eqref{Seq:poscond4} becomes
\begin{equation}
  \label{Seq:poscond4b}
  2D^{-2}-D^{-4}-
\left[ Q_1(0_{2l}\oplus\id_{2m-2l})Q_1^T \right]_{[2n,2n]}\geq 0
\end{equation}
where $[X]_{[2n,2n]}$ denotes the upper left $2n\times 2n$ block
  of $X$, $Q_1\in\mr{SO}(2m)$, and $l\leq m$ denotes as before the
    number of pure modes in $B'$. In contrast to \Eqref{Seq:poscond4},
  only the upper left $2n\times2n$ block of
  $Q_1(0_{2l}\oplus\id_{2(m-l)})Q_1^T$ enters the condition. That a
Choi rank of $\leq n$ (or $\leq m$) is \emph{not} 
  necessary for degradability in that case can be seen from a simple example: 
the $(n\to n+k)$-mode FGC $T:\gamma\mapsto (A\gamma A^T+B)\oplus
B_2$, where $T_{(A;B)}$ is a $n\to n$ degradable FGC and $B_2$ the CM of a
full-rank $k$-mode Gaussian state.  We can obviously degrade $T$ by discarding
the modes in state $B_2$, applying the degrading map (of the channel
$T_{A;B}$) to the remaining modes and adjoining a system in a pure
state that is a purification of $B_2$. On the other hand, the Choi
rank of $T$ is the Choi rank of $T_{A;B}$ plus $2k$, i.e., up to $m+k$.
\end{example}

\section{Extremality of fermionic Gaussian states}
We show now that Gaussian states are \emph{extremal} with respect to
certain entanglement measures among all states with the same covariance
matrix. This follows in the same way as for bosonic systems \cite{WGC06}
from the fact that FGS can be obtained via a non-commutative central limit and
will later be useful to compute the quantum capacity of certain FGCs.
\begin{theorem}[Extremality Theorem]
\label{Sth:extremality}
Let $ f:\mathcal{B(H}^{\otimes d})\to \mathbb{R}$ be a continuous functional,
which is strongly super-additive and invariant under local unitaries
[$f(U^{\otimes d}\rho U^{\dagger\otimes d})=f(\rho)$]. Then for every even
density operator $\rho$ describing $d$-partite fermionic system we have that
$f(\rho)\geq f(\rho_G)$, where $\rho_G$ is the even fermionic Gaussian state
with the same second moments as $\rho$.
\end{theorem}
\Proof.
 Let us start by giving the precise definition of strong super-
 and sub-additivity.
\begin{definition}[Strong super-additivity and strong sub-additivity]
\label{ssadddef}
 Let $\rho$ be a density operator on
 $\mathcal{H}:=(\mathcal{H}_{A_1}\otimes
 \mathcal{H}_{A_2})\otimes(\mathcal{H}_{B_1}\otimes
 \mathcal{H}_{B_2})$ and $\rho_i$, $i=1,2$ restrictions of $\rho$ on
 $\mathcal{H}_{A_i}\otimes\mathcal{H}_{B_i}$. Then the functional $f:
 \mathcal{B(H)}\to\mathbbm{R}$ is called strongly super-additive if for all $\rho$ it holds
\begin{equation}
f(\rho)\geq f(\rho_1)+ f(\rho_2)
\end{equation}
and equality holds for $\rho=\rho_1\otimes\rho_2$ (and with a
natural generalization to more parties).\\ 
Analogously, we can define so-called \emph{strong sub-additivity}, which
becomes useful when estimating channel capacities.
\end{definition}

Note that here we write the Hilbert space (Fock space) of $n+m$
  modes as the tensor product of $n$-mode Fock
  space and $m$-mode Fock space. For simplicity, we do not use the
  more precise notation $\otimes_f$ that would remind us that we're
  not dealing with the normal tensor product. However, we note here
  that as long as states and operations we consider respect the parity
  superselection rule, it holds (as in the standard case) that density
  matrices $\rho_1$ and $\rho_2$ representing fermionic 
states on disjoint sets of modes do commute (since fermionic states are even polynomials in the Majorana operators), justifying the notation
$\rho_1\otimes\rho_2$. Similarly, one can show that the ``local
fermionic operations'' ${\cal E}_i$ acting only on disjoint sets of modes
(and respecting the parity superselection rule) commute with each
other \cite{SSGK18}.

Let us consider a $d$-partite fermionic system in state $\rho$. Then
we denote for brevity's sake by $\rho^{\otimes n}$ the state of $n$
identical copies of the system in this state. With a slight abuse of
notation, we write $U^{\otimes d}\rho U^{\otimes d}$ to denote the
application of the local fermionic completely positive map that
corresponds to applying the same unitary $U$ to each of the $d$
subsystems corresponding to the $d$ parties considered. Then the basic
idea of the proof, as in bosonic case, is the following chain of
relations \cite{WGC06}:
\begin{align}
f(\rho)=\frac{1}{n}f(\rho^{\otimes n})&=\frac{1}{n}f(\underbrace{U^{\otimes d}\rho^{\otimes n} U^{\dagger\otimes d}}_{\tilde{\rho}})\label{extremunitary}\\ 
&\geq\frac{1}{n}\sum_{k=1}^n f(\tilde{\rho}_k)\rightarrow f(\rho_G)\label{extremlimit},
\end{align}
where the additivity, invariance under local unitaries (i.e., the
  identical unitary $U$ applied to all the modes (of the different
  copies of $\rho$) belonging to each of the $d$
  parties
and strong
super-additivity respectively is used (here $\tilde{\rho}_k$ denotes the
reduced state of the modes corresponding to the $k$th copy). The last
step is the result of the quantum mechanical central limit theorem for
anti-commuting variables \cite{Hud73}. 
 
Before proving the theorem we need to introduce the notation. As stated above, we
consider $n$ copies of a $d$ partite fermionic
system and denote the associated creation and annihilation operators
of the $i$th mode of the $j$th copy by 
$a^{j \dagger}_i$,
$a^{j}_i$ respectively, where $j=1, \dots, n$, $i=1,\dots, d$. Here
$a^{j \dagger}_i\equiv(a^{j}_i)^{\dagger}$. Note that for
  simplicity of notation we consider here a $d$-mode $d$-partite system,
  i.e., one mode per site. The extension to $L_j\geq1$ modes at site $j$ is
  straight forward (by introducing yet more indices: $a_{i,k}^{j}$, where index $k$ denotes $k$-th mode of the $j$-th copy of the $i$-th subsystem).
  
  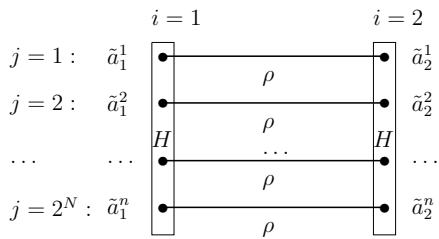
\begin{figure}[h]
  \centering
  \centerline{\large
  \resizebox{6.1cm}{!}{
    \begin{tikzpicture}[
      scale=1,
      duenn/.style={thin},
      rpfeil/.style={thick,->},
      lpfeil/.style={thick,<-},
    ]
    \draw (2.45,4) -- (2.85,4) -- (2.85,0.5) -- (2.45,0.5) -- (2.45,4);
    \draw (2.3,2.25) node [right] {$H$};
    \draw (6.45,4) -- (6.85,4) -- (6.85,0.5) -- (6.45,0.5) -- (6.45,4);
    \draw (6.3,2.25) node [right] {$H$};
    \draw[thick,-] (2.7,3.75) -- (6.6,3.75); 
    \draw  (6.4,3.725) node [right] {$\bullet$};
    \draw (2.4,3.725) node [right] {$\bullet$};
    \draw (4.3,3.35) node [right] {$\rho$};
    \draw[thick,-] (2.7,2.9) -- (6.6,2.9); 
    \draw  (6.4,2.875) node [right] {$\bullet$};
    \draw (2.4,2.875) node [right] {$\bullet$};
    \draw (4.3,2.5) node [right] {$\rho$};
    \draw (4.3,2) node [right] {$\dots$};
    \draw[thick,-] (2.7,1.85) -- (6.6,1.85); 
    \draw  (6.4,1.825) node [right] {$\bullet$};
    \draw (2.4,1.825) node [right] {$\bullet$};
    \draw (4.3,1.45) node [right] {$\rho$};
    \draw[thick,-] (2.7,1.0) -- (6.6,1.0); 
    \draw  (6.4,0.975) node [right] {$\bullet$};
    \draw (2.4,0.975) node [right] {$\bullet$};
    \draw (4.3,0.6) node [right] {$\rho$};
    \draw (2.3,4.45) node [right] {$i=1$};
    \draw (6.3,4.45) node [right] {$i=2$};
    \draw (-0.25,3.725) node [right] {$j=1:$};
    \draw (-0.25,2.875) node [right] {$j=2:$};
    \draw (-0.25,1.825) node [right] {$\dots$};
    \draw (-0.25,0.975) node [right] {$j=2^N:$};
    \draw (1.5,3.725) node [right] {$\tilde{a}^1_1$};
    \draw (1.5,2.875) node [right] {$\tilde{a}^2_1$};
    \draw (1.5,1.825) node [right] {$\dots$};
    \draw (1.5,0.975) node [right] {$\tilde{a}^n_1$};
    \draw (7.0,3.725) node [right] {$\tilde{a}^1_2$};
    \draw (7.0,2.875) node [right] {$\tilde{a}^2_2$};
    \draw (7.0,1.825) node [right] {$\dots$};
    \draw (7.0,0.975) node [right] {$\tilde{a}^n_2$};
    \end{tikzpicture}
  }
}

  \caption{The local ``Gaussification'' operation for a bipartite
    system. Given $n=2^N$ copies of a bipartite state, they are transformed by a
    local Gaussian unitary ($H$), which effects a basis change
    $a_i^j\to\tilde{a}_i^j$ such that the bipartite reduced state of
    the each 
    copy approximates the same bipartite Gaussian state the second
    moments of which are given by those of the initial (not necessarily Gaussian) state.
}
  \label{Sfig:Gaussification}
\end{figure}

The only step requiring a proof in \Eqref{extremlimit} is that
  $\frac{1}{n}\sum_{k=1}^n f(\tilde{\rho}_k)$ converges to $f(\rho_G)$
  where $\rho_G$ is the Gaussian state with the same second moments as
  $\rho$.  For a suitable choice of $U$, this follows along the lines
  of the central-limit theorem  proved in \cite{Hud73}. \\
To see this we let $n=2^N$ and choose $U$ as the local passive
Gaussian unitary, which transforms annihilation
operators at site $i$ as 
\[a_i^{j}\mapsto \tilde{a}_i^{j} = \sum_{l=1}^n
\frac{H_{jl}}{\sqrt{n}}a_i^{l},
\]  
where $H=\left( \begin{array}{cc} 1&1\\1&-1
\end{array} \right)^{\otimes N}$.  
We see that the first mode at site $i$ in the transformed system is now
described by the symmetric combinations
$\tilde{a}_i^1=\frac{1}{\sqrt{n}}(a_i^1+a_i^2+\dots a_i^n)$ of fermionic operators. 
In \cite{Hud73} it was shown that $\tilde{\rho}_1$ converges to
$\rho_G$ by showing that 
all cumulants except those of second order
(covariances) vanish and the latter coincide with the covariances of $\rho$. 
We need to generalize this result to the other, non-symmetric, combinations
appearing in modes 2 to $n$ at each site. 
The operators $\tilde{a}_i^j, j\geq2$ differ from $\tilde{a}_i^1$ only
by the appearance of minus signs to be applied to all operators
referring to one half of the copies involved. 
But since the different copies describe independent systems (as
evidenced by the tensor product structure of the state $\rho^{\otimes n}$)
and since for fermionic states only \emph{even} moments of the Fermi
operators are non-zero, these minus signs appear only in even powers
when computing the cumulants of $\tilde{\rho}_k, k\geq2$ and therefore
all the reduced states are the same and by \cite{Hud73} each of them
converges to the same Gaussian state $\rho_G$ as $n\to\infty$. $\qed$

The detailed convergence argument can be found in \cite{Greplova2013}.

There are several interesting functionals which satisfy the conditions
of the theorem, among them entropy, relative entropy, distillable
entanglement \cite{BDSW96}, and squashed entanglement \cite{ChWi03}.

%


\end{bibunit}

\end{document}